# Economic Implications of Corporate Governance and Corporate Social Responsibility: Evidence from Banks in Bangladesh

Liza Fahmida

## Abstract

This study explores the challenges and implications of Corporate Social Responsibility (CSR) in the banking sector of Bangladesh, highlighting its regulatory framework, implementation gaps, and alignment with sustainable development goals. While CSR is mandated by the central bank, the profit-driven nature of banking institutions often shifts the focus of CSR initiatives toward competitive advantage and brand enhancement rather than addressing genuine social and environmental needs. Major investments are concentrated in education and health sectors, with minimal attention to environmental sustainability and marginalized communities. Weak regulatory oversight, profit-oriented governance structures, and limited stakeholder participation hinder the effective implementation of CSR. The lack of diversity in board representation, particularly the exclusion of women and underrepresented groups, further limits the participatory and inclusive nature of CSR. This study underscores the need for stronger policy interventions, enhanced monitoring mechanisms, and a shift in corporate governance to transform CSR into a tool for meaningful societal impact. The findings call for further research to explore strategies for aligning profit-driven motives with sustainable and equitable development objectives.

## Bangladesh Economy

Bangladesh's economy has experienced steady growth since the early 2000s, achieving a GDP growth rate averaging around 6% per year by 2015. The country transitioned from being predominantly agrarian to having a more diversified economy, with significant contributions from the manufacturing and services sectors. The ready-made garments (RMG) industry emerged as a critical driver, accounting for over 80% of export earnings, alongside remittances from overseas workers playing a vital role in bolstering foreign exchange reserves.

Agriculture, while still important, saw its share in GDP decline as industrial and service sectors expand. Government efforts to reduce poverty resulted in a consistent decrease in the poverty rate, although challenges like income inequality, unemployment, and inadequate infrastructure persisted. By 2015, Bangladesh had positioned itself as a lower-middle-income country, as recognized by the World Bank, reflecting significant progress in socio-economic indicators.

## Bangladesh Banking Sector

The banking sector in Bangladesh is one of the key pillars of the country's economy. As of 2015, it comprised four major categories of institutions: state-owned commercial banks (SOCBs), private commercial banks (PCBs), foreign commercial banks (FCBs), and specialized banks focused on agriculture and industry.

SOCBs accounted for a substantial portion of total banking assets but faced challenges such as inefficiency, non-performing loans (NPLs), and political interference. In contrast, PCBs, including banks like Dutch-Bangla Bank Limited (DBBL) and Prime Bank Limited (PBL), demonstrated better governance and operational efficiency, leading to stronger financial performance and growth. FCBs had a limited presence but were known for their service quality and niche market focus.

By 2015, the sector had made strides in financial inclusion, with policies promoting microfinance and mobile banking services. However, systemic issues such as governance inefficiencies, high NPL ratios, and weak regulatory enforcement remained persistent challenges. Regulatory initiatives by the Bangladesh Bank, including mandatory CSR activities, were introduced to enhance accountability and promote sustainable practices across the sector. Despite these efforts, the sector's capacity to address socio-economic and environmental priorities required further improvements in governance and policy enforcement.

## Literature Review

The definition of Corporate Social Responsibility (CSR) has undergone significant evolution since the mid-20th century, resulting in diverse interpretations and connotations (Benn, 2011, p. 56). Despite this variability, CSR is widely recognized as the social and environmental accountability of businesses (Benn, 2011, p. 56). Initially rooted in philanthropy, CSR has transitioned toward an interactive and participatory approach to community development (Benn, 2011, p. 34). Carroll (2013, p. 30) described CSR as the societal impact of a company's actions, while also framing it as the duty of policymakers and decision-makers to safeguard and enhance societal welfare alongside their own interests (Carroll, 2013, p. 33).

Another perspective of CSR integrates its economic and legal roles within the broader responsibility framework, defining it as a commitment not only to comply with legal and economic obligations but also to foster societal welfare and sustainability (Carroll, 2013, p. 33). According to Allouche (2006, p. 68), CSR must reflect both economic efficiency and ethical principles in corporate behavior. Carroll's (1991, p. 40) four dimensions of CSR—economic, legal, ethical, and philanthropic—outline an organization's responsibilities toward achieving societal well-being and sustainable development. These elements highlight that businesses must generate profit, adhere to legal norms, maintain ethical standards, and contribute positively as corporate citizens (Carroll, 1991, p. 43). Expanding on this, Carroll (2010) emphasized that CSR extends beyond business operations to encompass legal, ethical, and discretionary expectations from society (cited in Benn, 2011, p. 56). This comprehensive view sees CSR as a multifaceted concept that involves participation from various stakeholders to regulate and mitigate the social and environmental effects of corporate activities, aiming to establish globally accepted business norms (Grosser, 2016, p. 66).

Garriga and Mele (2004) highlighted Frederick's (1994, 1998) observations on the evolution of CSR, identifying four distinct stages of its development (cited in Benn, 2011, p. 56). The first stage, CSR 1, examines the philosophical and ethical relationship between business and society. CSR 2 focuses on the institutional responses of businesses to societal pressures and environmental factors. CSR 3 takes a normative approach, emphasizing the integration of ethics and values in corporate practices. Finally, CSR 4 stresses fostering harmonious global relations by addressing the societal impact of managerial decisions (cited in Benn, 2011, p. 57).

Garriga and Mele (2004) also introduced a theoretical model of CSR based on Parsons' (1961) framework, which posits that any social system should encompass four key elements: environmental and resource factors, political dynamics, social integration, and cultural aspects (cited in Benn, 2011, p. 57). These components have informed the development of four distinct groups of CSR theories in the modern social context (cited in Benn, 2011, p. 57).

## Corporate Governance (CG) and CSR

The practice of Corporate Social Responsibility (CSR) is deeply intertwined with a company's governance decisions. Effective management must address social, ethical, and stakeholder issues to align with the organization's broader responsibilities (Carroll, 2013, p. 94). CSR encompasses a wide range of economic, political, environmental, cultural, market, and governance functions, making it a complex and multifaceted concept (Benn, 2011, p. 59). Decisions surrounding CSR are influenced by both the internal and external environments of organizations, incorporating commercial, environmental, social, and sustainability considerations (Benn, 2011, p. 61). This interconnection underscores the critical relationship between corporate governance (CG) and CSR practices.

Over the past decades, CG and CSR have become inseparable concerns for policymakers and governments, particularly in light of events such as the global financial crisis (Baldarelli, 2015, p. 6). Corporate governance serves as the foundational mechanism for organizational decision-making and operational control (Andrea, 2014, p. 2). Its primary purpose is to establish a framework for directing, managing, and administering an organization effectively (Carroll, 2013, p. 96). According to the OECD, CG entails the processes through which corporations are managed, the roles and responsibilities of managers and boards of directors, and the mechanisms for ensuring accountability to shareholders (Keong, 2002, p. 40).

Legitimacy, trust, and sound governance are crucial for an organization's long-term sustainability (Benn, 2011, p. 107). From a narrow perspective, CG focuses on protecting the interests of owners and shareholders, whereas a broader perspective encompasses the welfare of all stakeholders, including employees, suppliers, customers, and the wider community (Rao, 2016, p. 33). CSR is inherently connected to the governance structure of an organization, forming a framework for ensuring social and environmental accountability.

The European Commission emphasizes the adoption of an open governance framework to ensure CSR aligns with a global approach to sustainable development (EU Commission, 2001, 2002, 2006, cited in Baldarelli, 2015, p. 8). A positive relationship between CSR and CG is evident, as organizations with effective governance are generally more socially and environmentally responsible than those with weak governance practices (Chan, 2014, p. 68). CSR relies on robust stakeholder participation, which reinforces its alignment with good governance practices.

The banking sector exemplifies this dynamic, acting as a leader in corporate governance due to its extensive network of stakeholders, including customers, service providers, employees, regulators, and communities. This interconnectedness highlights the pivotal role of governance in shaping CSR practices within industries with a wide-ranging societal impact.

## CG in Banks

Banks and financial institutions bear significant accountability due to their reliance on depositors' money, which inherently ties their corporate governance (CG) to societal responsibilities. By providing critical financial services, banks have distinct governance requirements that set them apart from other organizations (Haan, 2016, p. 229). According to the Basel Committee guidelines, CG in banking encompasses the processes and conduct of boards of directors and executive management in overseeing investments, transactions, and daily operations. This governance framework is grounded in corporate objectives, accountability to shareholders and customers, and adherence to financial, operational, and regulatory safeguards. A key focus is on protecting depositors' interests while ensuring alignment between corporate activities and regulatory expectations (Zinca, 2012, pp. 279–280). These guidelines, referred to as "Enhancing CG for banking institutions," aim to provide a standardized approach to governance globally.

Common governance tools in organizations include board size and composition, ownership structure, management compensation, and market or corporate control mechanisms (Haan, 2016, p. 229). Shareholders appoint directors as a means of ensuring alignment between management actions and shareholder interests (Haan, 2016, p. 229). However, an oversized board may hinder value creation due to inefficiencies, such as the free-rider problem (Aebi et al., 2012, and Mehran et al., 2011, cited in Haan, 2016, p. 230). Independent directors must also avoid any connections with management to maintain governance integrity (Haan, 2016, p. 230). Concentrated ownership can strengthen management oversight, while compensation structures incentivize performance (Haan, 2016, pp. 230–232). Market controls like proxy contests, mergers, and hostile takeovers further ensure managers prioritize shareholder interests (Haan, 2016, p. 232).

In banking, these traditional governance tools differ significantly due to the unique risks and implications for economic stability. Banks depend on depositors for funding, and regulators monitor their activities on behalf of small depositors, making regulatory oversight a critical governance tool (Haan, 2016, p. 229). Key differences between banks and other firms include stricter regulations, distinct capital structures, and the complex nature of banking operations (Haan, 2016, p. 229). Banks play a pivotal role in economic activity and national prosperity due to their multidimensional influence on markets (Levine, 2004, p. 2).

The banking sector's sensitivity to financial, operational, and reputational risks underscores the importance of robust governance. Effective management and board oversight directly influence a bank's objectives, shareholder accountability, and depositor protections. The quality of governance varies by country, shaped by historical, economic, and cultural factors. Consequently, regulatory frameworks are essential external forces in shaping bank behavior, as they balance internal and external governance forces to align public and private interests optimally (Ciancanelli, 2000, pp. 1–24).

# CSR Practices in Bangladesh

Banks and financial institutions bear significant accountability due to their reliance on depositors' money, which inherently ties their corporate governance (CG) to societal responsibilities. By providing critical financial services, banks have distinct governance requirements that set them apart from other organizations (Haan, 2016, p. 229). According to the Basel Committee guidelines, CG in banking encompasses the processes and conduct of boards of directors and executive management in overseeing investments, transactions, and daily operations. This governance framework is grounded in corporate objectives, accountability to shareholders and customers, and adherence to financial, operational, and regulatory safeguards. A key focus is on protecting depositors' interests while ensuring alignment between corporate activities and regulatory expectations (Zinca, 2012, pp. 279–280). These guidelines, referred to as "Enhancing CG for banking institutions," aim to provide a standardized approach to governance globally.

Common governance tools in organizations include board size and composition, ownership structure, management compensation, and market or corporate control mechanisms (Haan, 2016, p. 229). Shareholders appoint directors as a means of ensuring alignment between management actions and shareholder interests (Haan, 2016, p. 229). However, an oversized board may hinder value creation due to inefficiencies, such as the free-rider problem (Aebi et al., 2012, and Mehran et al., 2011, cited in Haan, 2016, p. 230). Independent directors must also avoid any connections with management to maintain governance integrity (Haan, 2016, p. 230). Concentrated ownership can strengthen management oversight, while compensation structures incentivize performance (Haan, 2016, pp. 230–232). Market controls like proxy contests, mergers, and hostile takeovers further ensure managers prioritize shareholder interests (Haan, 2016, p. 232).

In banking, these traditional governance tools differ significantly due to the unique risks and implications for economic stability. Banks depend on depositors for funding, and regulators monitor their activities on behalf of small depositors, making regulatory oversight a critical governance tool (Haan, 2016, p. 229). Key differences between banks and other firms include stricter regulations, distinct capital structures, and the complex nature of banking operations (Haan, 2016, p. 229). Banks play a pivotal role in economic activity and national prosperity due to their multidimensional influence on markets (Levine, 2004, p. 2).

The banking sector's sensitivity to financial, operational, and reputational risks underscores the importance of robust governance. Effective management and board oversight directly influence a bank's objectives, shareholder accountability, and depositor protections. The quality of governance varies by country, shaped by historical, economic, and cultural factors. Consequently, regulatory frameworks are essential external forces in shaping bank behavior, as they balance internal and external governance forces to align public and private interests optimally (Ciancanelli, 2000, pp. 1–24).

## Data Analysis: Direct CSR Expenditure of DBBL and PBL

The trend of CSR expenditure shows that banks are focusing more on the education and health sectors and are making a very minimal contribution in the environment and other sectors. In the education sector, DBBL provides a major portion of their financing towards scholarships and stipends to poor and meritorious students, and, in the health sector, this bank has financed the Smile Brighter Program for cleft-lipped children, rural healthcare, and financing towards medical infrastructure (Dutch Bangla Bank Annual Report 2015, p.18). PBL provides finance to the Prime Minister's Relief Fund for the victims of natural calamity, annual stipends to poor students and has established an eye hospital and nursing institute (Prime BankAnnual Report 2015, p. 33).

Table 0-1: Direct CSR Expenditure: PBL and DBBL (in Million Taka)

| Particulars | Prime Bank Ltd. | | | | | Dutch-Bangla Bank Ltd. | | | | |
|---|---|---|---|---|---|---|---|---|---|---|
| | 2011 | 2012 | 2013 | 2014 | 2015 | 2011 | 2012 | 2013 | 2014 | 2015 |
| **Education** | 2.786 | 4.78 | 5.92 | 23.2 | 17.84 | 21.32 | 33.21 | 35.78 | 75.12 | 61.0 |
| **Health** | 4.14 | 8.915 | 3.73 | 4.45 | 1.51 | 2.76 | 3.5 | 5.17 | 83.12 | 19.47 |
| **Disaster Management** | 0.556 | 0.66 | 1.27 | 2.85 | 0.93 | 5.5 | 8.5 | 20.83 | 11.4 | 3.96 |
| **Environment** | 0 | 1 | 0 | 0 | 0 | 0.14 | 0.41 | 0.05 | 0.22 | 0.15 |
| **Sport** | 3.4 | 7.96 | 10.96 | 4.38 | 1 | 0.17 | 0.24 | 0.7 | 0.2 | 0 |
| **Art & Culture** | 0.935 | 2.3 | 0.49 | 0.17 | 0 | 0 | 2.48 | 0.75 | 4 | 0 |
| ***Cultural Welfare** | 0 | 0 | 0 | 0 | 0 | 0 | 0 | 0 | 0 | 0.45 |
| ***Infrastructure** | 0 | 0 | 0 | 0 | 0 | 0 | 0 | 0 | 0 | 2.6 |
| ***Income Generation** | 0 | 0 | 0 | 0 | 0 | 0 | 0 | 0 | 0 | 0 |
| **Others** | 5.392 | 3.95 | 3.35 | 3.35 | 2.48 | 4.2 | 15.24 | 15.24 | 7.3 | 2.5 |
| **Total** | 17.21 | 29.565 | 25.72 | 38.4 | 23.76 | 34.13 | 52.77 | 78.54 | 181.36 | 90.17 |

The CSR expenditures of Dutch-Bangla Bank Limited (DBBL) and Prime Bank Limited (PBL) from 2011 to 2015 reveal that education and health sectors were their primary focus, consistent with regulatory requirements mandating at least 30% of total CSR spending on education (Bangladesh Bank Circular, 2014, p. 2). These expenditures were used for initiatives such as stipends and scholarships. In 2015, both banks significantly reduced their overall CSR spending compared to 2014. Notably, DBBL allocated Tk. 83.12 million to the health sector in 2014, primarily for hospital construction, with approximately 97% of the health budget directed toward infrastructure development (Dutch-Bangla Bank Annual Report, 2014, p. 204).

In 2014, DBBL's total CSR spending reached Tk. 181.36 million but declined by nearly 50% in 2015 to Tk. 90.17 million. From 2011 to 2014, both banks did not invest in areas such as arts and culture, cultural welfare, infrastructure, or income generation. However, in 2015, DBBL allocated a small portion of its budget to cultural welfare and infrastructure. Additionally, DBBL launched awareness campaigns on social issues such as justice, drug addiction, and environmental protection, but these initiatives were more focused on branding and advertising rather than substantive community impact (Dutch-Bangla Bank Annual Report, 2015, p. 271).

Aside from education and health, disaster management emerged as the third largest area of CSR investment. However, environmental initiatives received minimal attention, with PBL making no contributions to the environment sector between 2013 and 2015. This lack of focus on environmental sustainability contrasts sharply with Bangladesh's constitutional commitment to sustainable development, as articulated in the United Nations Conference on Sustainable Development (UNCSD) charter of 2012 (National Report on Sustainable Development, 2012, p. 6).

The data suggest that banks in Bangladesh often use CSR as a strategic tool to enhance brand image rather than to address societal and environmental challenges meaningfully. This observation aligns with Barnea et al. (2010), who argue that organizations may increase CSR investments to secure reputational benefits for managers and major shareholders, rather than prioritizing societal impact (Barnea, 2010, pp. 71–84). While banks claim to pursue CSR as part of their corporate citizenship responsibilities (Prime Bank Annual Report, 2015, p. 33), their contributions have not demonstrated significant outcomes for broader societal welfare.

Simpson (2002) noted that effective CSR practices in the banking sector could create positive stakeholder relationships and enhance both social and financial performance (Simpson, 2002, pp. 106–107). However, neither DBBL nor PBL has shown evidence of fostering such relationships, despite receiving awards for their CSR activities. This disconnect may stem from insufficient monitoring and evaluation of CSR initiatives at the grassroots level.

The Bangladesh Bank (BB), as the regulatory authority, provides guidelines for CSR activities. These guidelines require banks and financial institutions to review CSR reports before allocating new funds (Bangladesh Bank CSR Report, 2014, p. 31). However, an examination of DBBL and PBL's annual reports reveals no clear evidence of systematic monitoring of their CSR initiatives.

Although BB established the Sustainable Finance Department in 2015 to integrate CSR and green banking practices into core business functions, there is limited published information about the evaluation and supervision of these initiatives (Bangladesh Bank Website, 2016). The absence of robust monitoring mechanisms raises questions about the effectiveness of CSR implementation in achieving meaningful societal and environmental benefits.

The mandatory nature of CSR for banks in Bangladesh, enforced through central bank regulations, has sparked debates about whether CSR should be voluntary or compulsory (Horrigan, 2010, p. 25). Given the profit-oriented nature of banking institutions, there are doubts about their willingness to invest in CSR solely for societal and environmental benefits. Effective CSR requires coordinated governance, active civil society participation, rigorous regulatory oversight, and a clear understanding of societal needs. However, gaps in governance practices, stakeholder engagement, and regulatory enforcement reflect a lack of alignment in assessing and addressing CSR priorities in Bangladesh.

The CSR expenditures of banks in Bangladesh, while substantial in education and health sectors, show limited focus on environmental sustainability and societal impact. Without effective monitoring and a shift toward more inclusive governance, CSR activities risk becoming mere tools for corporate branding rather than drivers of sustainable development. Addressing these challenges requires enhanced regulatory frameworks, stronger stakeholder participation, and a commitment to aligning business practices with national and global sustainability goals.

## Economic Implications of CSR in Bangladesh

Corporate Social Responsibility (CSR) in Bangladesh has gained momentum as an essential tool for fostering economic development, especially in underprivileged sectors. The banking industry, led by institutions such as Dutch-Bangla Bank Limited (DBBL) and Prime Bank Limited (PBL), has played a pivotal role in integrating CSR into their operational strategies. This section explores the broader economic implications of these activities, highlighting trends, challenges, and potential growth opportunities.

Both DBBL and PBL have consistently directed significant portions of their CSR expenditure toward education and health. For instance, between 2011 and 2015, DBBL allocated over 30% of its total CSR expenditure to education, funding scholarships for underprivileged students and supporting healthcare initiatives like rural medical infrastructure. PBL, while operating on a smaller scale, also invested in similar areas, albeit with a lesser financial commitment. These contributions have tangibly impacted access to basic services for marginalized communities, facilitating skill development and reducing long-term poverty.

By investing in infrastructure such as hospitals and educational institutions, banks indirectly contribute to job creation and local economic activity. For example, DBBL's investment in healthcare infrastructure in 2014 saw a substantial expenditure of Tk. 83.12 million, creating a ripple effect in related industries.

CSR activities also enhance the brand value and reputation of banks, which translates into better customer loyalty and competitive positioning. Both DBBL and PBL have leveraged their CSR initiatives to strengthen their market presence, as reflected in the steady growth of their net asset values (NAV) during the study period.

**Statistical Analysis and Trends**

DBBL displayed a sharp rise in CSR spending, peaking at Tk. 181.36 million in 2014, before declining to Tk. 90.17 million in 2015. PBL's spending peaked in 2014 at Tk. 38.4 million but remained significantly lower compared to DBBL. The majority of expenditures for both banks were concentrated in education and health, while other areas like environmental sustainability received negligible funding. Education and health combined accounted for over 70% of DBBL's CSR budget, reflecting a focus on high-impact areas. However, critical sectors like environmental conservation and income generation received less than 5% of total allocations, raising concerns about balanced economic development.

Figure 1: Trends in CSR Expenditure on Education and Health (2011-2015)

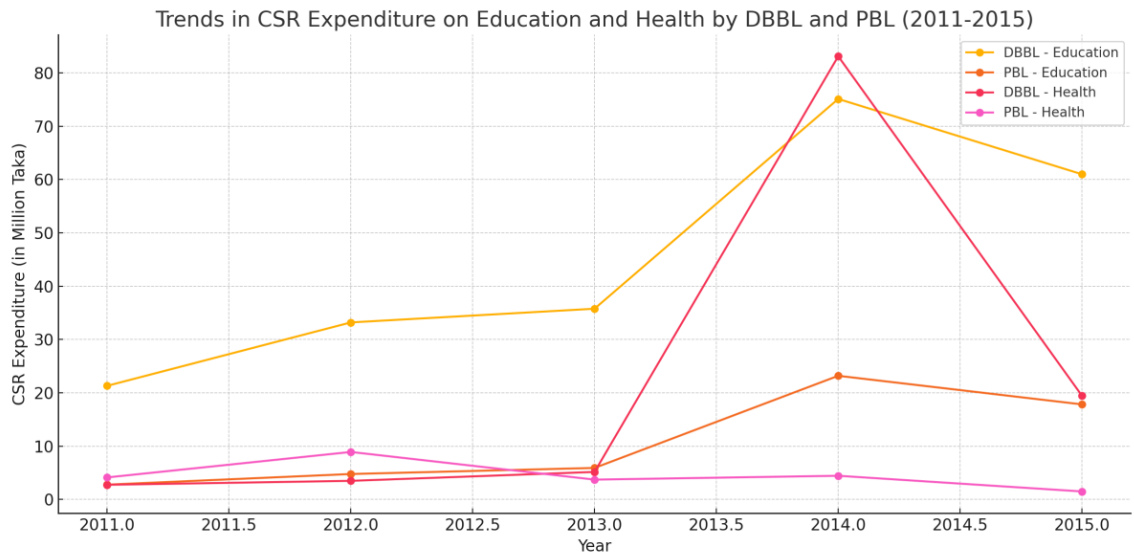

Figure 2: Total CSR Expenditure Trends by DBBL and PBL (2011-2015)

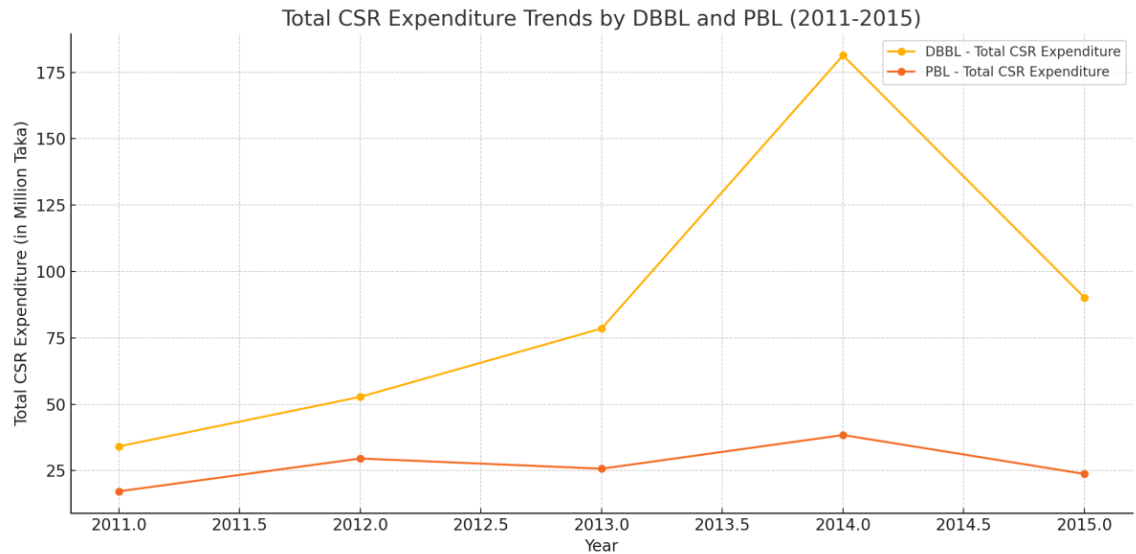

## Challenges and Areas for Improvement

Despite mandatory CSR guidelines by the Bangladesh Bank, there is limited evidence of robust monitoring and evaluation mechanisms. The absence of transparent reporting on outcomes undermines the potential for sustained impact.

Investments in environmental initiatives have been minimal. Given Bangladesh's vulnerability to climate change, banks need to prioritize funding for green projects and renewable energy programs.

The governance structures of banks often lack representation from marginalized groups, including women and indigenous communities. This exclusion hinders the participatory nature of CSR initiatives and reduces their socio-economic reach.

## Conclusion

Undertaking CSR initiatives poses challenges for banking institutions, as their primary focus is on profit generation. One of the main criticisms of CSR is that the funds allocated for social development ultimately belong to shareholders, raising questions about whether such expenditures align with their interests (Horrigan, 2010, p. 24). If organizations fail to adopt CSR as an ethical commitment to sustainable development for future generations, their social, moral, and corporate responsibilities will remain under scrutiny. A key debate revolves around whether CSR should be voluntary or enforced through regulation (Horrigan, 2010, p. 25). In Bangladesh, CSR activities by banks and financial institutions are mandated by the central bank. However, there are notable gaps in addressing genuine social needs. Many banks appear to prioritize CSR spending as a means to achieve competitive advantages and enhance growth and profitability. This situation is exacerbated by weak regulatory oversight, profit-driven governance structures, and limited stakeholder participation in decision-making processes. These factors highlight the need for stronger policy interventions to transform CSR into a meaningful and effective economic tool. Additionally, the lack of representation of marginalized groups and women on bank boards further undermines efforts to adopt inclusive and participatory CSR practices. Finally, the profit-centric nature of the banking sector raises questions about whether mandatory CSR can achieve its intended objectives. Further research is needed to explore how CSR can be successfully implemented in such a context to drive sustainable and equitable development.